\documentclass[times,twocolumn,final, nopreprintline]{elsarticle}

\usepackage{framed,multirow}

\usepackage{appendix}

\usepackage{amssymb}
\usepackage{amsmath}
\usepackage{latexsym}

\usepackage{url}
\usepackage[dvipsnames]{xcolor}

\usepackage{booktabs}
\newcommand{\ra}[1]{\renewcommand{\arraystretch}{#1}}

\usepackage[ruled, vlined]{algorithm2e}

\usepackage{soul}

\usepackage[percent]{overpic}
\usepackage{subfigure}

\usepackage{hyperref}
\hypersetup{
    colorlinks=true,
    linkcolor=blue,
    filecolor=magenta,      
    urlcolor=cyan,
}
\usepackage{cleveref}

\usepackage[switch,pagewise]{lineno} 

\def\R{\mathbb{R}}
\renewcommand{\O}[1]{\mathcal{O}({#1})}

\newcommand{\frep}{p}
\newcommand{\geom}{\mathbf{g}}
\newcommand{\generator}{\mathcal{G}}
\newcommand{\baryspace}{\Delta}
\newcommand{\gb}{\beta}
\newcommand{\bv}{\mathbf{v}}
\newcommand{\be}{\mathbf{e}}
\newcommand{\nor}{\mathbf{n}}
\newcommand{\dst}{\mathbf{d}}
\newcommand{\smbox}{\footnotesize{\Box}}
\newcommand{\imagsmbox}{\footnotesize{\Diamond}}
\newcommand{\ms}{micro-structure}
\newcommand{\Ms}{Micro-structure}
\newcommand{\powpces}{\nu}
\newcommand{\dbox}{box}
\newcommand{\ibox}{cuboid}
\newcommand{\iboxes}{cuboids}
\newcommand{\imagTet}{S}

\newcommand{\figref}[1]{Figure~\ref{#1}}
\newcommand{\secref}[1]{Section~\ref{#1}}
\newcommand{\algref}[1]{Algorithm~\ref{#1}}

\definecolor{deemph}{gray}{0.0}
\newcommand{\deemph}[1]{\textcolor{deemph}{#1}}

\journal{Computers \& Graphics}

\begin{document}


\begin{frontmatter}

\title{Plane-Activated Mapped Microstructure}

\author[1]{Jeremy Youngquist\corref{cor1}}
\cortext[cor1]{Corresponding author: }
\emailauthor{jyoungquist@ufl.edu}{Jeremy Youngquist}
    
\author[2]{J{\"o}rg Peters}

\author[2]{Meera Sitharam}

\fntext[fn1]{University of Florida}


\begin{abstract}
Querying and interacting with models of massive material micro-structure requires localized on-demand generation of the micro-structure since the full-scale storing and retrieving  is cost prohibitive.
When the \ms\ is efficiently represented as the image of a canonical structure under a non-linear space deformation to allow it to conform to curved shape, the additional challenge is to relate the query of the mapped \ms\ back to its canonical structure.

This paper presents an efficient algorithm to pull back a mapped \ms\ to a partition of
the canonical domain structure into \dbox es and only activates \dbox es whose image is likely intersected by a plane. 
The active \dbox es are organized into a forest whose trees are traversed depth first to generate mapped \ms\ only of the active \dbox es. The traversal supports, for example, 3D print slice generation  in  additive manufacturing.
\end{abstract}


\end{frontmatter}




\begin{figure}[h]
    \centering
    \includegraphics[scale = 0.36]{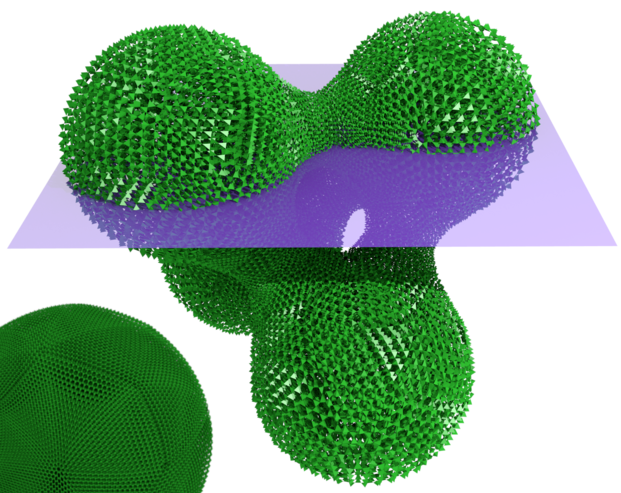}
    \caption{Mapped \ms\ and slice plane. The map is piecewise cubic.  {\it Left foreground:} High-resolution \ms.}
    \label{fig::costonbubblemesh}
\end{figure}

\section{Introduction}
\label{sec::intro}

State-of-the-art 3D printers are capable of printing structure at the micrometer \cite{Fogel2018_MetallicMicrostructures} and even nanometer \cite{Vyatskikh2018_NanoArchitectedMetals} scale. 
Additive manufacturing therefore, in principle, allows designs that spatially vary and optimize the material properties of objects via their \ms\ \cite{Wang18_GradedLattice, Bernd10_GoalBased, Schumacher15_ControlElasticity, Panetta15_ElasticTextures}.
However, a one meter cube with micrometer structures challenges existing storage capacities and deliberate access. 
The natural response, to generate the \ms\ on demand, is not without its own problems.
New fast printers are able to print fine-scale \ms s over regions up to one million times larger than the feature size \cite{Saha19_ScalableSubmicrometer, Jonusauskas19_Mesoscale3DPrint}.
Therefore \ms\ generation must be very fast and activated with keen focus on the query.

An additional challenge is to conform to curved shapes without breaking the structure at a boundary, compromising its structural integrity. 
One solution is to generate \ms\ as the mapped image of a canonical structure, i.e. under a non-linear space deforming map. 
This kind of 3D morphing  allows the \ms\ to fill,
for example, a curved tetrahedral partition of a macro-shape as in \figref{fig::costonbubblemesh} adapted from \cite{CoST_SPM19,Feng18_CurvedDelaunay}.
However, any query now has to be pulled back from the mapped image to the canonical domain \ms.

This paper presents an efficient algorithm to pull back and so activate mapped micro-structure intersected by a plane; and so enable slice generation of massive mapped \ms s for 3D printing.
The approach generates \ms\ only in a close neighborhood of the plane in $\R^3$, and is agnostic as to the shape of the canonical structure 
and its partition into local pre-images.
We will illustrate the approach by hex-paving the tetrahedral domain of a piecewise total degree 3 map in Bernstein-B\'ezier form.

\begin{figure*}[h!]
    \centering
    \begin{overpic}[scale = 0.48]{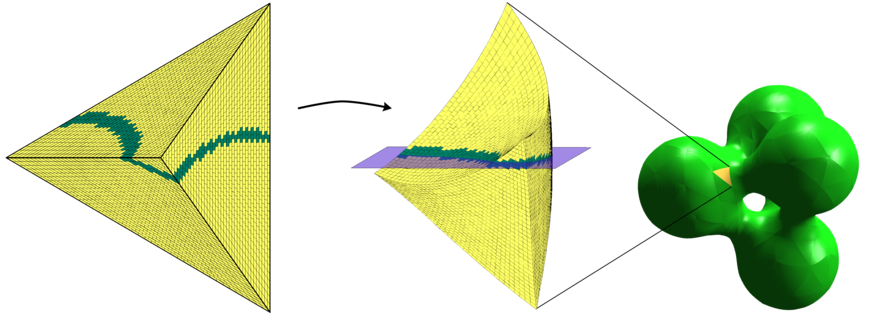}
        \put(38, 26){\huge$\displaystyle\geom$}
    \end{overpic}
    \caption{Mapping $\geom(\smbox_\gb)$.
    {\it Left}: Tetrahedron $\baryspace$ with \dbox\ partition.
    {\it Middle}: Curved $\geom(\baryspace)$ with overlapping \iboxes.
    Note that the pre-image of the slicing plane is a curved surface.  {\it Right}: The full set $\geom_i(\baryspace)$ with 
    the exterior face of the specific $\geom(\baryspace)$ is marked yellow.
    }
    \label{fig::preimage}
\end{figure*}

\section{Related Work}
\label{sec::relatedwork}

\subsection{Slicing}
\label{sec::relatedwork::slicing}

Surface slicing algorithms take as input a triangulated surface and return a set of oriented boundary curves representing the intersection of that surface with a set of slice planes.
These boundary curves are used to generate infill and machine instructions for the printer.

Leveraging massively parallel architecture, GPU-based methods can provide up to a 30x speedup over similar CPU-based algorithms \cite{Wang17_GpuSlicingContinuousPrint, Zhang17_GpuSlicing, Dant2016_OptimalMeshSlicingParallel} and out-of-core slicing algorithms reduce the memory burden for larger meshes by keeping the mesh on the hard drive rather than the memory of the computer\cite{Choi02_TolerantSlicing, Vatani09_SlicingNearestNeighbor}.
However, these out-of-core methods have  quadratic complexity in the number of triangles.
An optimal slicing algorithm  \cite{Minetto2017_OptTriSlice}
can be designed to have  complexity 
linear in the number of triangles,
the number of slice planes, and 
the number of triangle-plane intersections,
provided the slices have uniform thickness.
However, this algorithm takes as input the entire mesh, and thus is ruled out for procedurally generated or massive micro-structures.

Generating micro-structure on-demand has been explored \cite{Vidimce13_OpenFab}. 
However the materials are defined directly as the voxelized signed-distance map in the physical space $\R^3$ rather than as a parametric map of some simpler pre-image. It therefore does not conform to curved shapes and does not follow the spirit of state-of-the-art modelling and analysis tools, such as isogeometric analysis.
Isogeometric analysis uses parametric maps, splines or B\'ezier functions.
A second drawback of the approach in \cite{Vidimce13_OpenFab} is that each slice has to be generated in its entirety and is therefore limited in size and fineness of the \ms. 

In this paper, we address both challenges, introducing a streaming architecture capable of printing \ms s defined on parametric spaces with slices of massive size.

\subsection{Procedural Microstructure}
\label{sec::relatedwork::micro}

Procedural \ms s are defined by a function $\frep : \R^3 \rightarrow \{0, 1\}$ identifying points belonging to the \ms\ \cite{Pasko95_FRepModeling, Pasko11_ProceduralMicrostructures, Martinez16_ProceduralVoronoiFoams}.
In order to be computationally efficient, $\frep$ must have $\O{1}$ time and space complexity independent of the evaluation point.
Furthermore, if the structure is to be printed, $\frep$ must define support structures (if required) and satisfy overhang criteria \cite{Martinez16_ProceduralVoronoiFoams}.

Many classes of sufficiently fine-scale \ms s behave as if they were a continuum, with the continuum behavior increasing in accuracy as the \ms\ grows finer \cite{OstojaStarzewski07_RVEBook, Ostoja-starzewski_latticemodels, Huet90_Variational}.
This allows for specification of functionally graded material properties of procedural models while still treating them with continuum analysis.
This has been successfully integrated into analysis tools to optimize microstructure to conform to a spatially varying compliance matrix \cite{Wang18_GradedLattice, Bernd10_GoalBased, Schumacher15_ControlElasticity, Panetta15_ElasticTextures}.

However, for such \ms\ to be used with state-of-the-art isogeometric analysis tools, they should be defined in the parametric space of the isogeometric elements.  
Defining \ms\ in the parametric space of a free-form deformation is also  necessary to leverage the standard tools of computer-aided geometric design and conform the \ms\ to curved objects \cite{CoST_SPM19, Gupta19_ProgrammedLattice, Elber19_MicroTiles}.
Although \ms\ defined by $\frep$ directly in $\R^3$ is mostly straightforward to locally generate on-demand \cite{Vidimce13_OpenFab}, for $\frep$ mapped by a macro-scale parametric function 
it is necessary to find the pre-image of every voxel in $\R^3$ -- and isogeometric elements do not in general have analytic inverses.

The algorithm presented in the following
addresses these challenges by generating on-demand \ms\ defined on parametric elements.


\newcommand{\uu}{\mathbf{u}}
\newcommand{\ga}{\alpha}
\newcommand{\dg}{m}
\section{Preliminaries and Definitions}
\label{sec::defs}

We consider a typical 3D deformation map \cite{Sederber86_FreeForm}
in \emph{total degree} $\dg$
Bernstein-B\'ezier form (BB-form),
see e.g.\ \cite{Farin01_CurvesSurfaces}:
\begin{align}
   \geom &: \baryspace\subset \R^3 \rightarrow \R^3
   \qquad
   \sum\ga_i=\dg,\ 
   \ga_i\ge 0,
   \notag 
   \\
   \geom(\uu) &:=
   \sum_{\ga}\geom_{\ga} B_{\ga}(\uu) ,
   \quad
   B_{\ga}(\uu) := \frac{\dg!}{\ga_0!\ga_1!\ga_2!\ga_3!} 
   \prod_{i=0}^{3}
   \uu^{\ga_i}_i,
   \label{eq:BBform}
\end{align}
Since injectivity is mandatory when generating \ms\ and large deformations unduly stretch or squeeze, we may assume that $\geom$ satisfies
$0 < \det{\nabla\geom} < M$ for some constant $M$.
That is the Jacobian determinant is nonzero and bounded. 
Assuming, without loss of generality, a slicing plane of constant $z=z_0$
and denoting the  z-coordinate of $\geom$
by $\geom^z$, the Pre-image Theorem 
then certifies the pre-image  $\{ \uu: \geom^z(\uu) = z_0 \}$ is bivariate without jumps, and with holes only due to the boundaries of $\baryspace$.

\subsection{Safe pre-image traversal}
The key to efficiently activating polyhedral sub-regions $\smbox_\gb$ of $\baryspace$ in order to fill it with micro-structure is to determine whether 
$\geom(\smbox_\gb)$ intersects the slice plane.  
For the vertices $\bv_i$  of $\smbox_\gb$, let $\imagsmbox_\gb$ be a polyhedral sub-region of $\R^3$ delineated by $\geom(\bv_i)$.
We will test against an enlargement of $\imagsmbox_\gb$ whose intersection with the slicing plane then identifies  $\smbox_\gb$ with $\geom(\smbox_\gb)$ that can intersect the plane.

The enlargement  depends on how much
$\geom(\baryspace)$ differs from the
linear function $\ell: \baryspace\to\R^3$ that interpolates
the mapped vertices $\geom(\bv_i) = \geom_{3\be_k}$ that form a simplex $\imagTet$.
Generalizing an estimate for the  bivariate total degree case \cite{Filip86}, we obtain for the trivariate case
\begin{align}
\sup_{\uu\in \baryspace} \| \geom - \ell \| \le \frac{1}{8}
\sum_{i,j\in \{1,2,3\}} 
  l_il_j \sup_{\uu\in \baryspace} \| \partial_{\uu_i\uu_j} \geom \|
\end{align}
where $l_i$ is initially the unit length of the domain with respect to variable $\uu_i$.
Here the mixed partials $i\ne j$ are counted twice
for a total of 9 terms of which only 6 are different.

For $\geom$ of degree 3, $\partial_{\uu_i\uu_j} \geom$ is linear, and each of the six expressions, attains its maximum at one of its four vertices.
As is well-known for polynomials in BB-form \cite{farin-curves-and-book-88}, the values of $\partial_{\uu_i\uu_j} \geom$ at each of  its four vertices are $6$ times the second differences at the vertex.
Equivalently, we compute for $k \in \{1,2,3,4\}$  
the six second differences of the BB-coefficients
\begin{align}
\dst_{ijk} :=
\geom_{3\be_k}-\geom_{2\be_k+\be_i}
-\geom_{2\be_k+\be_j}+\geom_{\be_k+\be_i+\be_j},
\quad
\text{ for } i\ne k\ne j
\end{align}
Then $\geom(\baryspace)$ is enclosed by offsetting $\imagTet$
in $x$  by 
\begin{align}
\mu^x := 
\frac{6}{8}
\sum_{i,j\in \{1,2,3\}} 
  l_il_j \max_{k} |\dst^x_{ijk}|
  \label{eq:bound}
\end{align}
where $\dst^x_{ijk}$ is the $x$-coordinate of $\dst_{ijk}$,
and offset likewise in $y$ and $z$.

If we split each edge of the domain $\baryspace$ into $n=2^\powpces$ pieces
then $l_i l_j$ becomes $4^{-\powpces}$ times the initial unit edge length and for  polyhedral $\smbox_\gb$  with side-lengths $l_i<2^{-\powpces}$, it suffices to test  $\imagsmbox_\gb$ offset by $\mu/4^\powpces$.
The algorithm intersects the edges of $\imagsmbox_\gb$ with the  planar slice 
using a tolerance of $\mu/4^\powpces$.

While tighter bounds of the images of the $\smbox_\gb$ can be obtained, by explicit subdivision of $\geom$  or by sleves \cite{Peters:2004:SLE}, we expect the number of sub-regions to be very large so that  $4^{-\powpces}$ is the dominant factor.

%
%
\subsection{Box paving}

To emphasize the generality of the approach, we choose the sub-regions $\smbox_\gb$ of $\baryspace$ to be deformed cubes, called \dbox es.  Then the $\imagsmbox_\gb \in \R^3$ are called (image) $\iboxes$.
Crucially, the \dbox es will conform to and fill the domain $\baryspace$ without overlap. So we do not need to be concerned about partially filled or clipped \dbox es as occur when superimposing a uniform grid  on $\baryspace$. 
By contrast, the \iboxes\ $\imagsmbox_\gb$ typically overlap when offset by $\mu$, but this is of little consequence since
they only serve to identify their domain \dbox es.

\def\wid{0.8\linewidth}
\def\wids{0.35\linewidth}
\begin{figure}[h]
    \centering
     \begin{overpic}[scale = 0.35, trim={40 60 200 70}, clip, width=\wid]{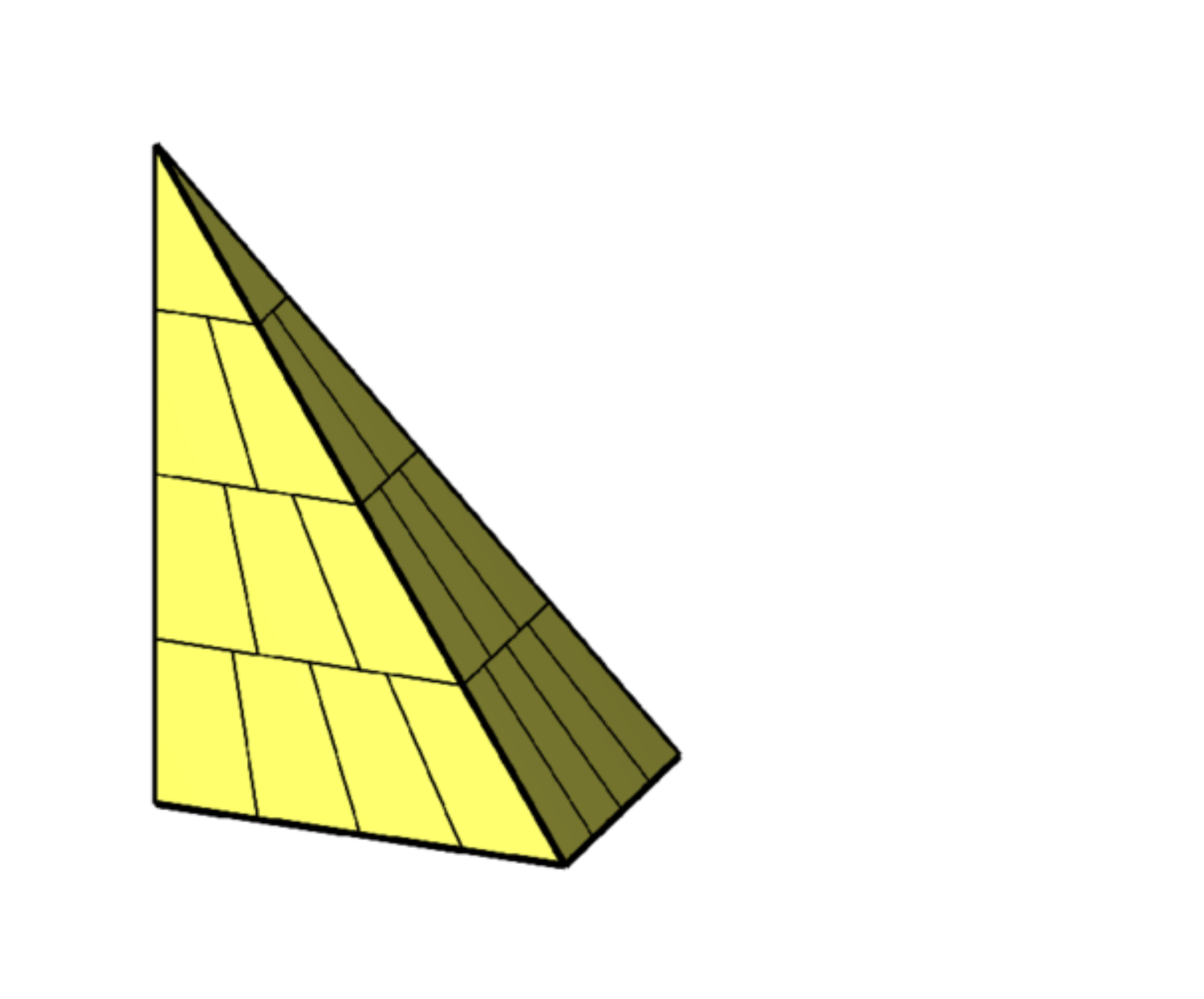}
        \put(50, 38){\footnotesize{ijk}}
        \put(35, 40){\footnotesize{120}}
        \put(23, 40){\footnotesize{110}}
        \put(13, 40){\footnotesize{100}}
        \put(12, 15){\footnotesize{000}}
        \put(25, 15){\footnotesize{010}}
        \put(37, 15){\footnotesize{020}}
        \put(47, 15){\footnotesize{030}}
        \put(12, 80){\footnotesize{300}}
    \end{overpic}
    \caption{Paving of a tetrahedral $\baryspace$ by \dbox es  with resolution $n=2^\powpces$. 
    }
    \label{fig::boxexample}
\end{figure}

\figref{fig::boxexample} illustrates the approach by paving a tetrahedron so that layer $i$ has $n-i$ rows and row $j$ of the layer $i$ has $n - i - j$ \dbox es.
Each \dbox\ can then be identified by the ID triple of integers $(layer, row, col)$.
At the apex, the paving degenerates into
a tetrahedron but this is not a problem since
\dbox es are only used to identify neighbors in $\baryspace$.
Each \dbox\ has up to 12 neighbors in this partition.

While, at first, it may seem odd to pave a tetrahedron with \dbox es, we note that a partition into small tetrahedra requires more complex indexing to visit neighbors.

\def\wid{0.7\linewidth}
\begin{figure}[h]
    \centering
    \includegraphics[width=\wid]{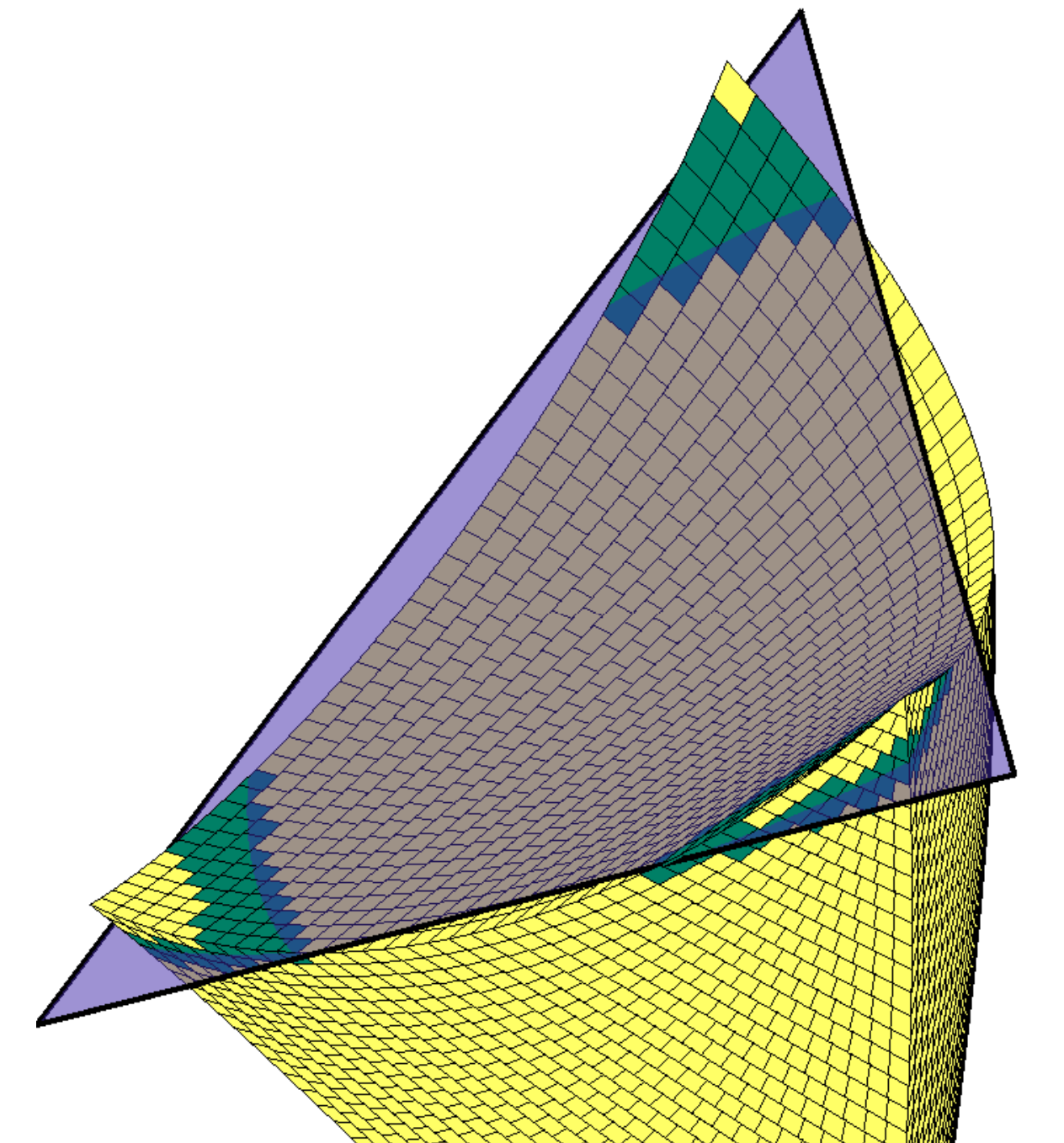}
    \includegraphics[width=\wid]{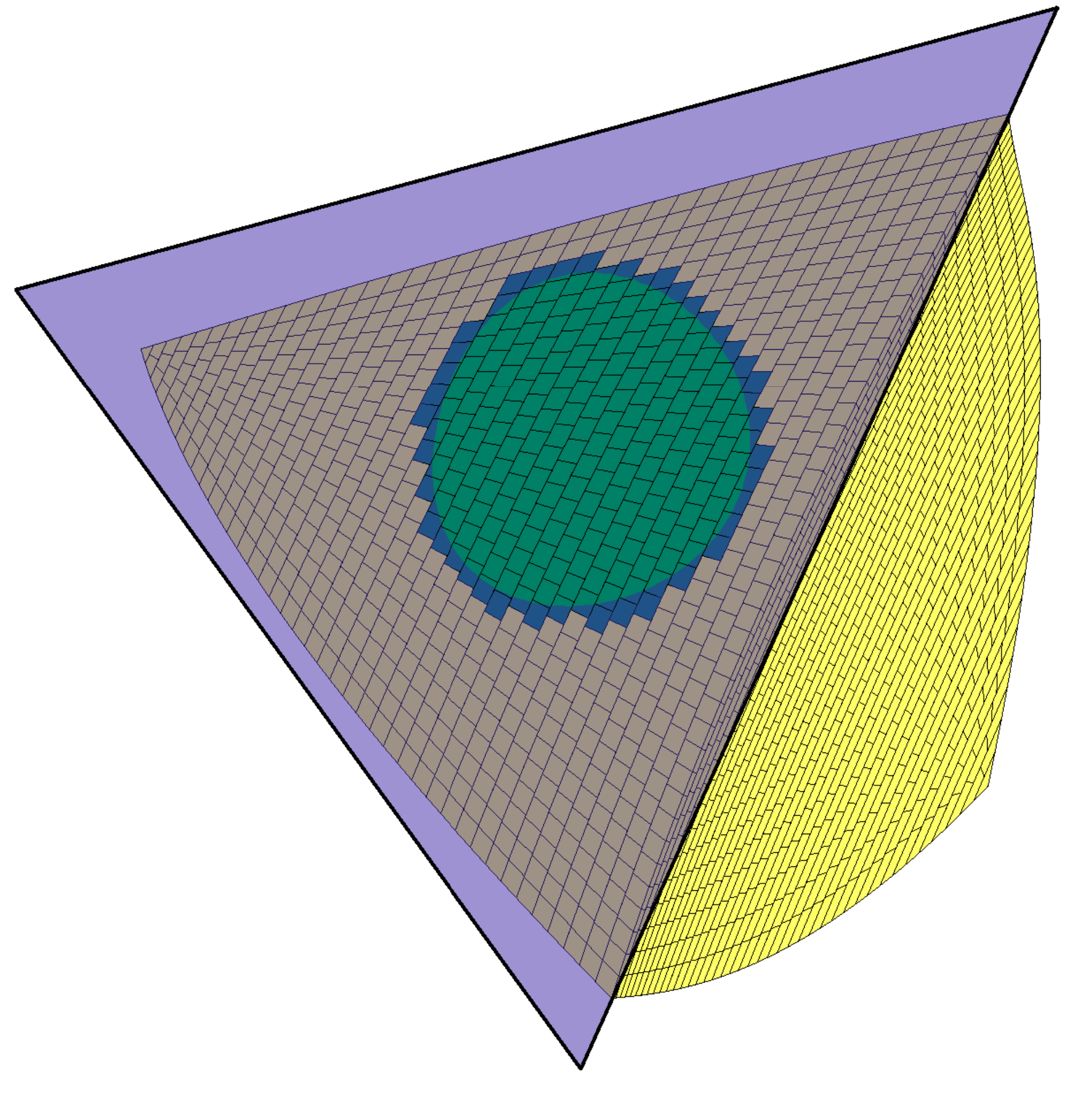}
    \caption{
    \textit{Top}: A single slice with multiple components due to intersecting edges.
    \textit{Bottom}: Closed loop intersection on a face (does not intersect any edges).
    }
    \label{fig::closedloop}
    \label{fig::multiplecomponents}
\end{figure}


\section{Algorithm}
\label{sec::algorithm}
Given a set $G$ of maps $\geom_i$, a partition of the shared domain $\baryspace$ into  \dbox es $\smbox_\gb$, and a set of planes $p_j \in P$, the goal is to efficiently traverse,
for each $\geom_i$  and $p_j$, all \dbox es $\smbox_\gb$ such that  $\geom_i(\smbox_\gb) \cap p_j \not = \emptyset$.
The key to efficiency is to activate (generate or retrieve from temporary storage) only
those \dbox es whose image can overlap the current slice plane $p_j$.
Then depth-first traversal minimizes generation and temporary storage of \ms.

\subsection{Reduction to a single map and plane}
\begin{figure}
    \centering
    \includegraphics[scale = 0.5]{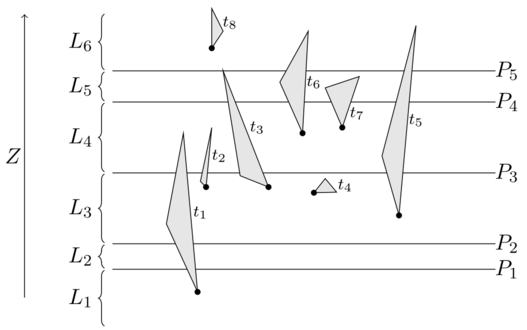}
    \caption{Illustration of the tetrahedral list algorithm, from \cite{Minetto2017_OptTriSlice}.  Each $L[i]$ contains the tetrahedra whose lowest value is between $P[i-1]$ and $P[i]$.}
    \label{fig::trianglelist}
\end{figure}

After sorting the $\geom_j$ into range list $L$ (see \figref{fig::trianglelist}) 
the outer loop of the algorithm performs an optimized plane sweep through all $\geom_j$ that possibly intersect plane
$p_i$.  Appendix A lists the simple main \algref{alg::main}
where highlighted routines perform the intersection of 
one map-plane pair discussed below and  $L[i]$ in  \algref{alg::iter::buildtetlist} is the set of $\geom_j$ which become active at plane $P[i]$. 
The 
minimal $z$ coordinate of the B\'ezier coefficients of $\geom_j$,
 $\geom^z_j$ 
must lie between the  $z$ coordinates of $p_{i}$ and $p_{i+1}$:
$P[i] \leq \geom^z_j< P[i+1]$.
The convex hull property of the BB-form then guarantees that no $\geom_j$ outside the list is intersected by the slicing plane.
Building the list, \algref{alg::iter::buildtetlist}, generalizes \cite{Minetto2017_OptTriSlice} to maps $\geom_j$; this can be optimized if the $\geom_j$ are already sorted or the planes have uniform thickness.

In the following we  illustrate the algorithm with a single polynomial map $\geom$ of total degree 3 in Bernstein-Bezier form, $\baryspace$ a tetrahedron, and a single plane $p$ with normal $(0, 0, 1)$.

\subsection{Components of one map-plane intersection} 
Due to the curvature of $\geom$, 
 the intersection $\geom(\baryspace)\cap p$  
can  have multiple disconnected components
(see \figref{fig::multiplecomponents}).
If we consider each \dbox\ as a node connected via an edge to its neighboring \dbox es,
the algorithm traces out a tree for each component starting a depth-first search from a \dbox\ known to straddle the plane.
Since we can assume that the pre-images of slices through $\geom(\baryspace)$
are surfaces, disconnected components occur only by slicing through the boundary.
We find all disconnected components by testing the \iboxes\ 
of the edges of $\imagTet$ (\algref{alg::iter::findboundaryboxes}),  and then test the faces of $\imagTet$ for closed loops.
A closed loop intersection can occur only if a face's
surface $\geom^s(t_1,t_2)$, $i=0,1,2,3$ has a normal $\nor^s$ orthogonal to the slicing plane \cite{Sinha85_Intersection}.
If the BB-coefficients of 
$\det (\nor, \partial_{t_1} \geom^s,  \partial_{t_2} \geom^s)$
are of one sign there is no loop. 
If the criterion fails to rule out an intersection, we find any 
by traversing and testing all \iboxes\ of the face.

\begin{figure}[h]
    \centering
    \includegraphics[scale = 0.50]{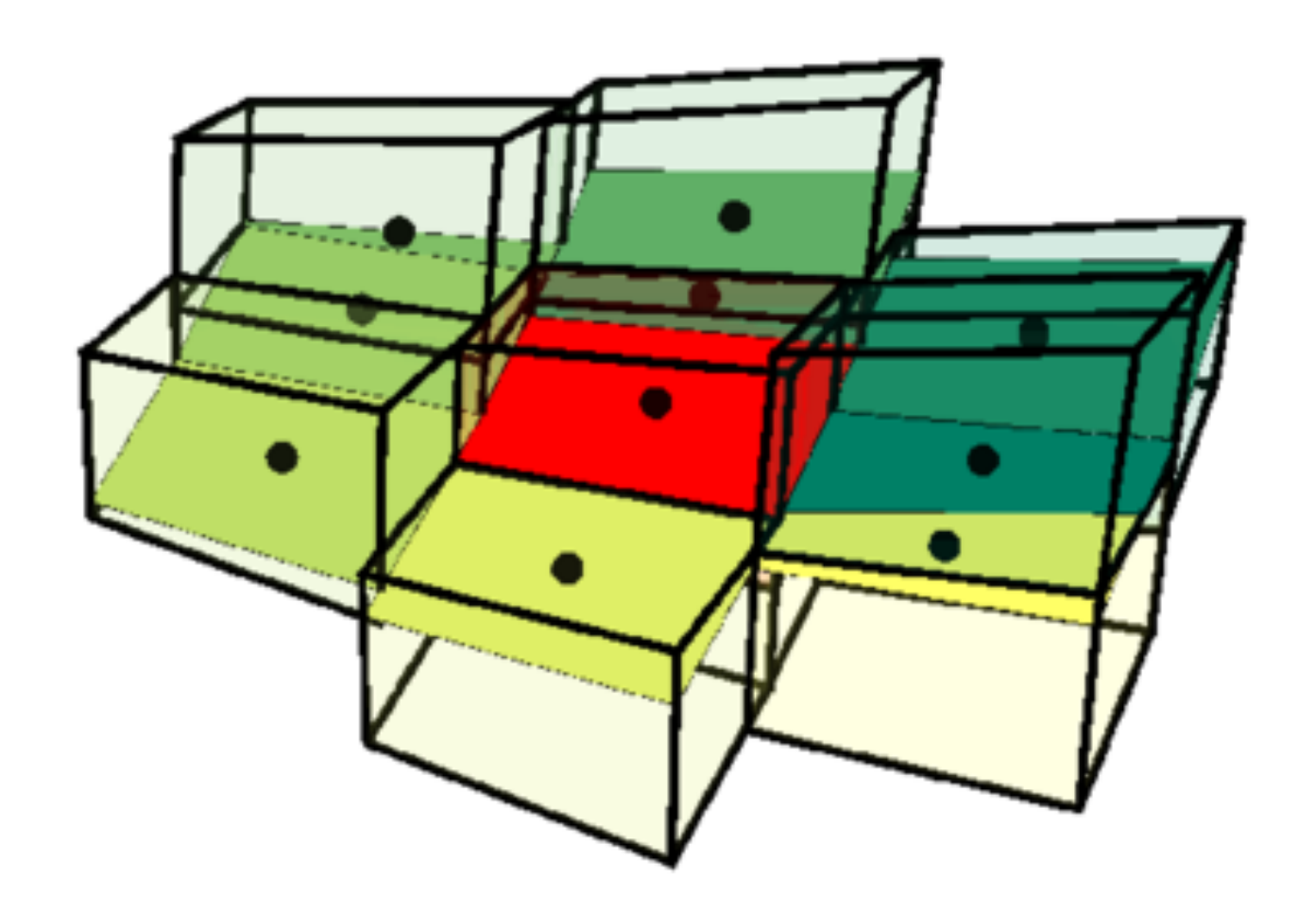}
    \includegraphics[scale = 0.44]{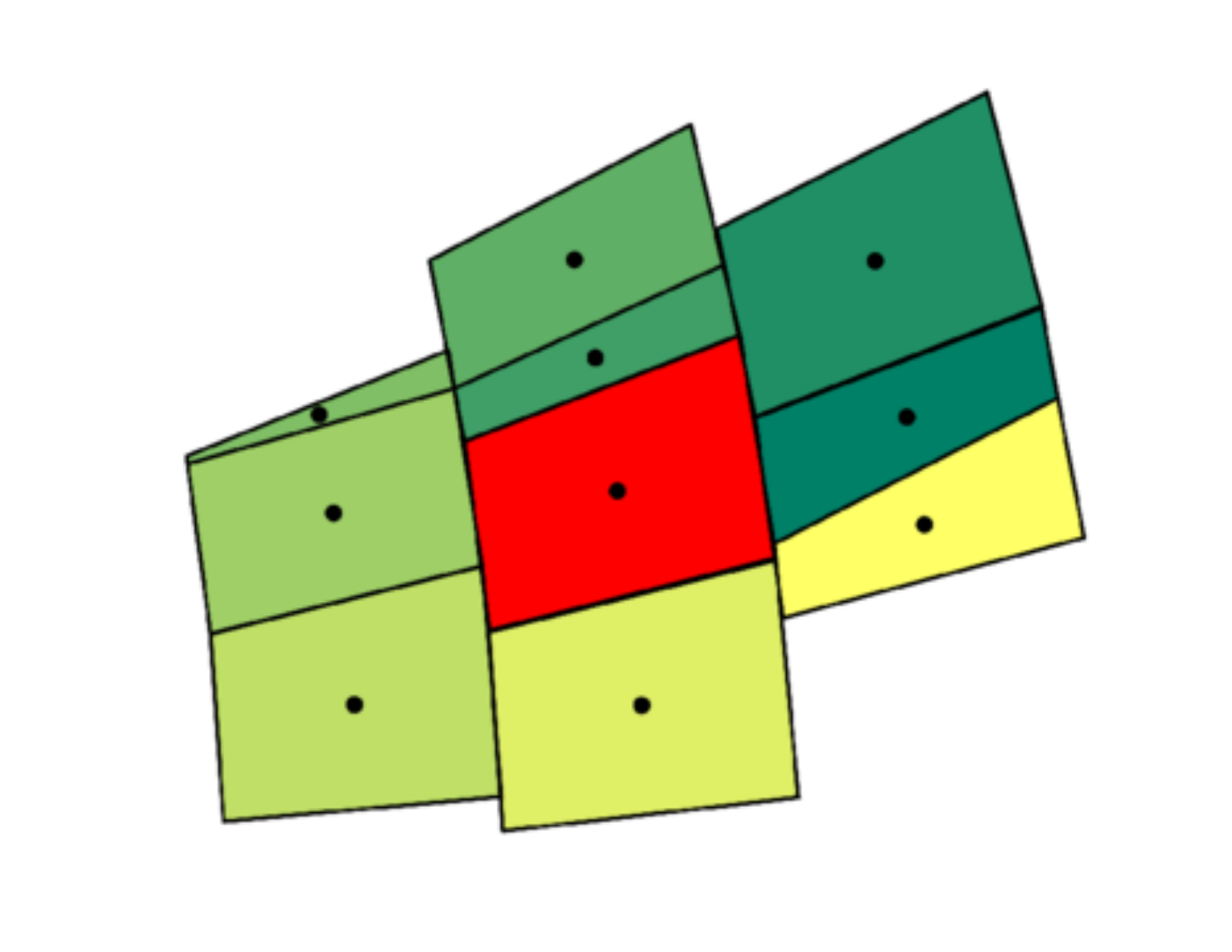}
    \caption{Planar cut through active \iboxes\ $\imagsmbox_\gb$. {\it Top:} 3D view, {\it bottom:} intersection with slice plane only. The current $\imagsmbox_\gb$ is red, the neighbors are sorted counterclockwise by their centers $\bullet$ and
    colored in traversal order from  dark green to yellow. 
    }
    \label{fig::boxbary}
\end{figure}

\subsection{Traversal of one map-plane pair}

For each new tetrahedron-plane pair, the algorithm
initializes (\algref{alg::interloop} in Appendix B) and starts with a \dbox\ found by \algref{alg::iter::findboundaryboxes}.
A stack, $toRevisit$, 
and a hashmap, $visitedBoxes$, 
store the \dbox\ IDs of  \ibox-plane intersections.

With each iterator increment,
we are either at the first and only \dbox\ of an isolated component and 
jump to another component by \algref{alg::iter::restart}; 
or 
we activate an intersecting neighbor (\algref{alg::iter::findnext}).

\subsection{Iteration}

\begin{figure}[h]
	\centering
	\includegraphics[width=0.35\textwidth]{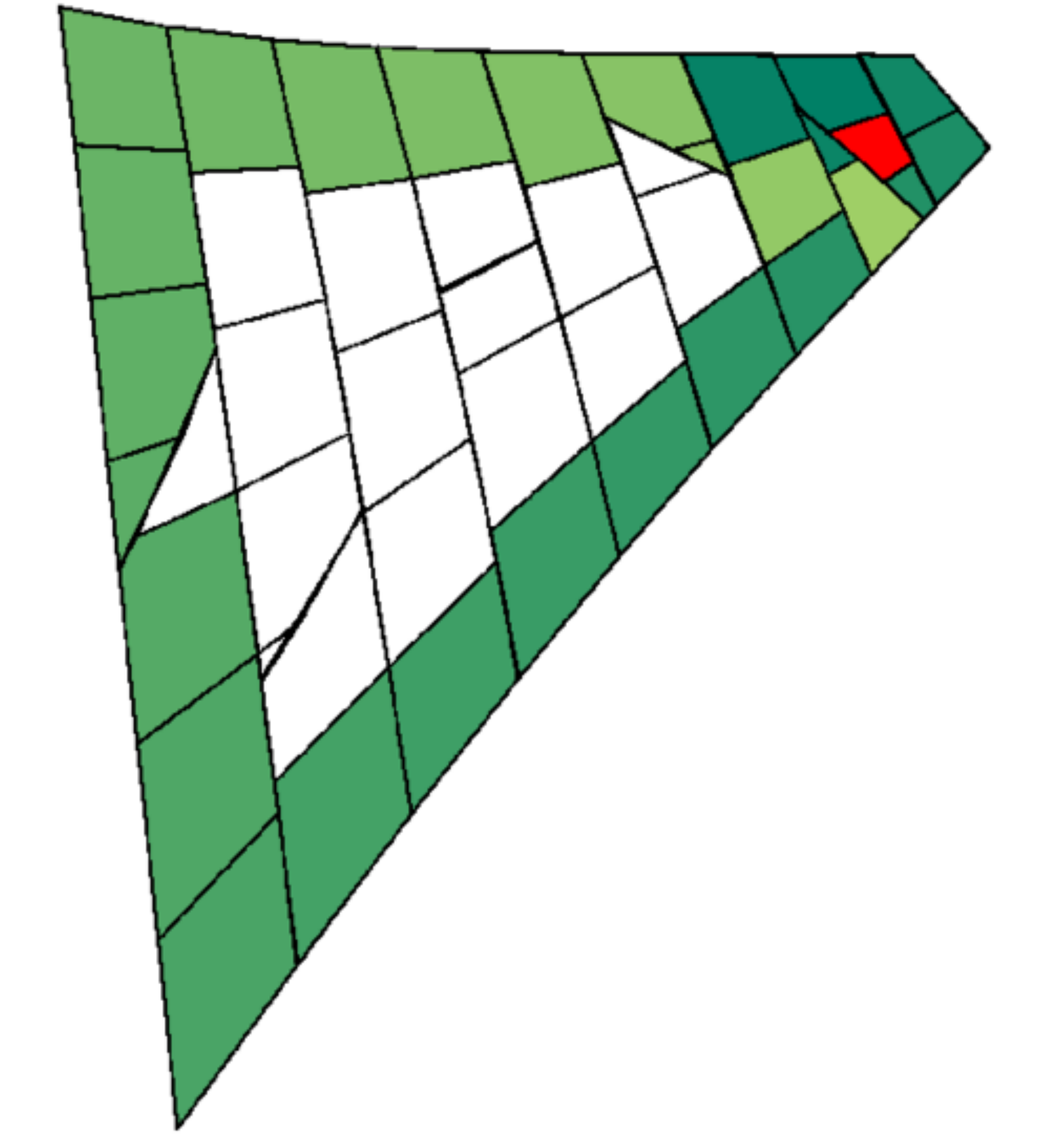}
	\caption{
	A red leaf node of the traversal tree:  all neighbors have been visited, but the slice is not yet fully covered and requires backtracking. 
	The completed traversal tree is  shown in \figref{fig::treepath}.
	}
    \label{fig::stuckinhole}
\end{figure}

To find the next \dbox\ (\algref{alg::iter::findnext}) we first find the set of neighboring \dbox es whose \ibox s intersect the plane and,
associating each \dbox\ with the center of its \ibox's intersection with the plane (see $\bullet$ in \figref{fig::boxbary}),
sort the centers clockwise about the center of the current \dbox\ with respect to the previous \dbox.
After removing the \dbox es we have already visited,
we proceed to the next \dbox\ in the list, setting $currBox$ to this \dbox\ ID and adding the ID to both $toRevisit$ and $visitedBoxes$.
Once all eligible neighbors have been visited, \figref{fig::stuckinhole},
we backtrack (\algref{alg::iter::walkback}) by popping \dbox\ IDs off
the $toRevisit$ stack until an eligible \dbox\ is found.

If the $toRevisit$ stack is empty,  we continue with  another connected component (\algref{alg::iter::restart}), 
restarting the iteration with a \dbox\ ID in 
the list of eligible \emph{boundary} \dbox es until
empty. 
If there are no more \emph{boundary} \dbox es, then the intersection is complete and $isValid$ is set to false.

\begin{figure}[h]
    \centering
    \includegraphics[scale = 0.3]{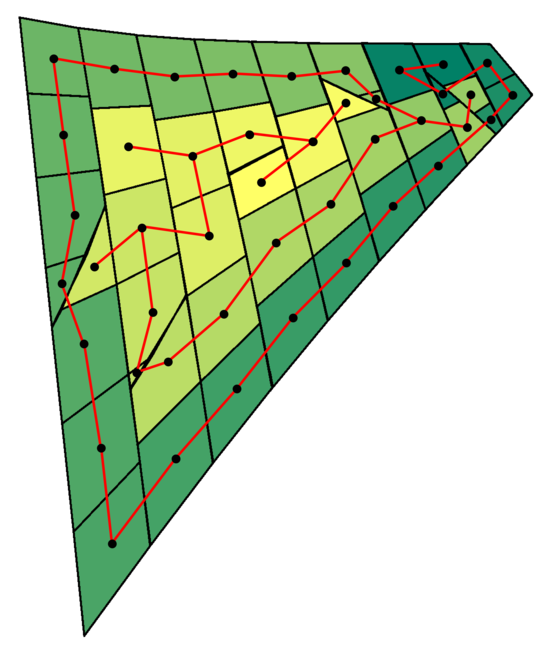}
    \caption{The path traced out by the algorithm is a tree.}
    \label{fig::treepath}
\end{figure}


\section{Complexity Analysis}
\label{sec::analysis}

\begin{figure*}[h!]
    \centering
    \includegraphics[scale = 0.5]{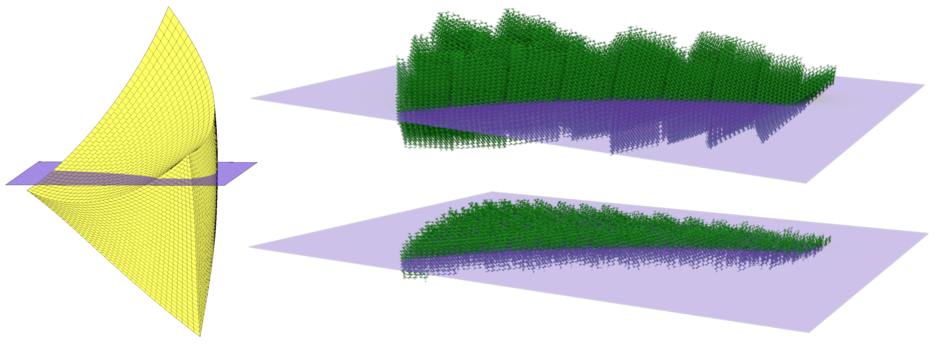}
    \caption{As the number of \dbox es increases, less \ms\
    ({\it bottom vs. top}) needs to be generated.}
    \label{fig::boxesvsmicrostructure}
\end{figure*}

Here we analyze the complexity of activating a single map-plane pair.

Letting $n$ be the number of \dbox es along one axis of $\baryspace$ there are $\O{n^2}$ \dbox es on each face of the tetrahedron.
Because the pre-image of the slice plane is a bivariate manifold, there are $k{n^2}$ $\imagsmbox_\gb$ straddling the slice,
where $k$ depends on the tightness of the estimate $\mu^x$ of \eqref{eq:bound} since
the \iboxes\ will overlap.

Each push or pop on the $toRevisit$ stack, and each get or put on the $alreadyVisited$ hashmap has unit cost.

\subsection{Initialization}

The complexity of initializing the iterator is bounded by the time it takes to build the list of \iboxes\ that straddle the boundary.
It takes just $\O{n}$ to iterate across each edge but $\O{n^2}$ to iterate over a face. 

\subsection{Iteration}

Construct a graph by treating each intersecting \ibox\ as a node connected to its neighboring \iboxes\ by edges.
Since there are $\O{n^2}$ \ibox s in the intersection, this graph has $\O{n^2}$ vertices.
The algorithm traces out a spanning forest of this graph, which is linear in the number of vertices.
Thus the iteration complexity is $\O{n^2}$.

\subsection{Space Complexity}

The $\O{n^2}$ triple integer lists of IDs, $toRevisit$ and $visitedBoxes$, are the only data structures that grow in size proportional to $n^2$.  
They are lightweight and reset for each tetrahedron and slice plane.


\section{Generating \Ms}
Since our focus is on accessing the \ms, \secref{sec::algorithm} intentionally did not discuss the actual slicing of the \ms\ with the mapped \dbox\ for streamed printing, which depends on the details of
the printing setup and \ms\ definition.

For a fixed \ms, increasing the number $n$ of \dbox es decreases the \dbox\ size and hence the amount $g(n)$
of \ms\ to be generated per \dbox.
Thus the total complexity for iterating through the slice and generating the \ms\ is $\O{n^2 g(n) + n^2}$.
Since $g(n)$ is inversely proportional to $n$, note that $\O{n^2 g(n)}$ can decrease with $n$ due to reduced thickness, see \figref{fig::boxesvsmicrostructure}.

Trading time and space complexity, we can re-generate \ms\ repeatedly for each slice, or store it under its \dbox\ ID in a list of active \dbox es as the plane moves.


\paragraph{Examples}

\begin{figure}[h]
    \centering
    \includegraphics[width=0.45\textwidth]{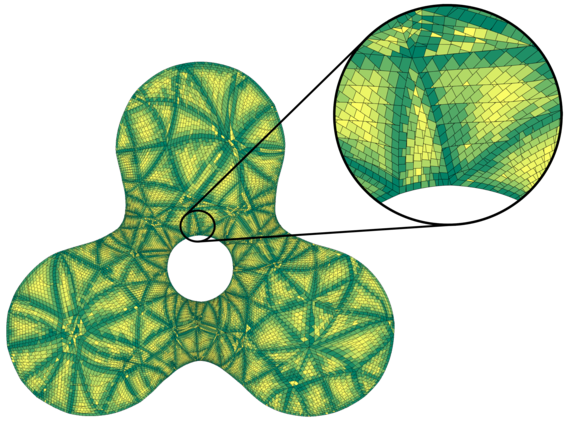}
    \caption{A slice through the mapped \ms\ of \figref{fig::costonbubblemesh}.  Each tile records one of 24,789 intersections between a \ibox\ and the slice plane. The colors represent the ordering of \dbox es within each map.
    }
    \label{fig::entirebubblemesh}
\end{figure}

The practicality of the algorithm is illustrated by slicing an extremely fine \ms\ on a mesh with 738 maps, 235 of which intersect the slice plane.
\figref{fig::entirebubblemesh} shows the slices of the \iboxes\ colored for traversal per $\geom_i$ from dark to light.

\begin{figure}[h]
    \centering
    \includegraphics[width=0.45\textwidth]{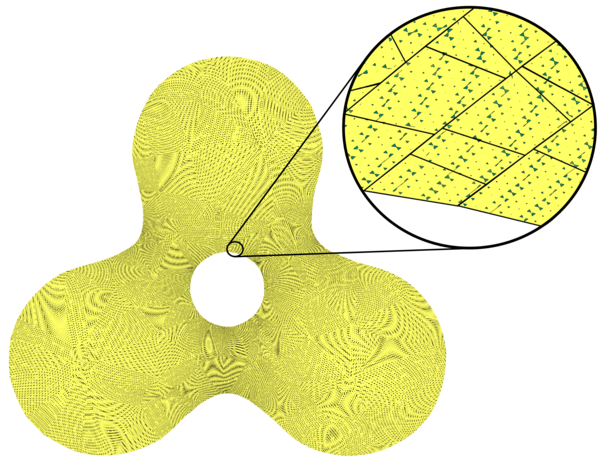}
    \caption{Slice through the mapped \ms\ of \figref{fig::costonbubblemesh}.  
    The enlargement reveals the \ms\ and \ibox s.}
    \label{fig::microstructurezoom}
\end{figure}

\figref{fig::microstructurezoom} shows the \ms\ generated on the mesh.

Table \ref{table::runtimes} lists run times for the generation of \dbox es, not including \ms\ generation or slicing.

\begin{table}[]
\ra{1.3}
\begin{tabular}{@{}lllll@{}} \toprule
n   & Time(s)  & \begin{tabular}[c]{@{}l@{}}Boxes in \\ Intersection\end{tabular} & \begin{tabular}[c]{@{}l@{}}Total Boxes \\ in Tetrahedron\end{tabular} & Intersect/Total \\ \midrule
4   & 1.797e-4 & 7                     & 20                 & 35.00\%      \\
8   & 6.368e-4 & 22                    & 120                & 18.33\%      \\
16  & 2.729e-3 & 79                    & 816                & 9.681\%      \\
32  & 1.410e-2 & 312                   & 5984               & 5.214\%      \\
64  & 0.1037   & 1,256                 & 45,760             & 2.745\%      \\
128 & 1.462    & 4,914                 & 357,760            & 1.374\%      \\
256 & 21.59    & 18,826                & 2,829,056          & 0.6655\%     \\
512 & 405.1    & 76,793                & 22,500,864         & 0.3413\%     \\
\bottomrule
\end{tabular}
\label{table::runtimes}
\caption{Run times and number of \dbox es $\smbox_\gb$ in the intersection {\it vs} number in the entire $\baryspace$.}
\end{table}

\section{Conclusion}
\label{sec::conclusion}

This paper proposed and implemented an algorithm for selectively generating mapped \ms\ only near a plane of interest.
The algorithm partitions parametric space into  \dbox es and uses their neighborhood relations to generate only \ms\ contained in those \dbox es whose images intersect the plane. The algorithm 
has optimal asymptotic complexity.

We applied the algorithm to the slicing problem for additive manufacturing, demonstrating that our algorithm is capable of generating large slices of extremely fine-scale \ms.
The implementation 
can easily be applied to hexahedra $\baryspace$ in place of tetrahedra.


\bibliographystyle{cag-num-names}
\bibliography{refs}


\newpage

\section*{Appendix A: Pseudocode: Slicing Multiple $\baryspace_i$}
\label{sec::app:buildtetlist}
\begin{algorithm}
\caption{Repeated Slicing Loop}
\label{alg::main}
\SetAlgoLined

    \SetKwInOut{Input}{input}
    \Input{$n, T, k, P, \generator$}
    \deemph{$L[1,\cdots, k+1] \gets$ \textsc{buildTetrahedronList}$(n, T, k, P)$\\}
    \deemph{$A \gets \{ \}$ set of active tetrahedra \\} 
    \For{\deemph{$ i \in {1, \cdots, k}$}}{
        \deemph{ $A \gets A \cup L[i]$\\}
        \For{\deemph{$tet \in A$}}{
            \eIf{\deemph{$tet.z_{max} < P[i]$}}{
                \deemph{$A \gets A \setminus {tet}$\\}
            }{
                \hl{\textsc{Iterator-Initialize}$(tet, P[i])$\\}
                \While{\hl{$\textsc{Iterator}-isValid$}}{
                    \hl{$M \gets tet.\geom(\generator(\textsc{Iterator-}currBox))$\\}
                    \deemph{\textsc{SliceAndPrint}$(M, P[i])$\\}
                    \hl{\textsc{Iterator-Increment}$()$\\}
                }
            }
        }
    }
    \Return

\end{algorithm}

\begin{algorithm}
\caption{BuildTetrahedronList}
\label{alg::iter::buildtetlist}
\SetAlgoLined

    \SetKwInOut{Input}{input}
    \SetKwInOut{Output}{output}
    \Input{$n, T, k, P$}
    \Output{$L$}
    
    $L[1, \cdots, k+1] \gets \{\}$\\
    \For{each $tet \in T$}{
        $i \gets$ min $i$ such that $P[i] \geq tet.z_{min}$\\
        $L[i] \gets L[i] \cup {tet}$\\
    }
    
    \Return{$L$}

\end{algorithm}

%
%
\section*{Appendix B: Pseudocode: Single Map-Plane \Ms\ Activation}

\begin{algorithm}
\caption{Single Map-Plane loop}
\label{alg::interloop}
\SetAlgoLined

    \SetKwInOut{Input}{input}
    \Input{$tet, plane, \generator$}
    \textsc{Iterator-Initialize}$(tet, plane)$\\
    \While{$\textsc{Iterator}-isValid$}{
        $M \gets tet.\geom(\generator(\textsc{Iterator-}currBox))$\\
        \deemph{\textsc{SliceAndPrint}$(M, plane)$\\}
        \textsc{Iterator-Increment}$()$\\
    }
    \Return

\end{algorithm}


\begin{algorithm}
\caption{Iterator-FindBoundaryBoxes}
\label{alg::iter::findboundaryboxes}
\SetAlgoLined

    \SetKwInOut{Input}{input}
    \Input{$tet$}
    $boundBoxes \gets \{ \}$\\
    \For{each \dbox\ on an edge of $tet$}{
        \If{$tet.\geom($box$)$ intersects $P$}{
            $boundBoxes \gets boundBoxes \cup box$\\
        }
    }
    \For{each face $F$ of $tet$}{
        \If{$F.N(u,v)$ is parallel to $P.normal$ for any $u, v$}{
            \For{each \dbox\ on $F$}{
                \If{$tet.\geom($box$)$ intersects $P$}{
                    $boundBoxes \gets boundBoxes \cup box$\\
                }
            }
        }
    }
    \Return $boundaryBoxes$ \\

\end{algorithm}

\begin{algorithm}
\caption{Iterator-Increment}
\label{alg::iter::increment}
\SetAlgoLined

    $candidates = intersectingNeighbors(currBox, T, P)$\\
    \eIf{$candidates$ is not empty}{
        \textsc{FindNextBox()}  \tcp*[r]{Typical case}
    }{
        \textsc{restartOnBoundaryBox()} \tcp*[r]{Only one \dbox\ in component}
    }
    \Return

\end{algorithm}




\begin{algorithm}
\caption{Iterator-FindNextBox}
\label{alg::iter::findnext}
\SetAlgoLined

    $candidates \gets intersectingNeighbors(currBox, T, P)$\\
    $sortCCW(candidates, currBox, prevBox)$\\
    $removeVisitedBoxes(candidates, visitedBoxes)$\\
    \eIf{$candidates$ is not empty}{
        $prevBox = currBox$\\
        $currBox = candidates[0]$\\
        $visitedBoxes.insert(currBox)$\\
        $toRevisit.push(currBox)$\\
    }{
        \textsc{WalkBack()} \tcp*[r]{At a leaf \dbox}
    }
    \Return

\end{algorithm}

\begin{algorithm}
\caption{Iterator-WalkBack}
\label{alg::iter::walkback}
\SetAlgoLined

    \eIf{$toRevisit$ is not empty}{
        \While{$toRevisit.top() == currBox$}{
            $toRevisit.pop()$\\
        }
        $prevBox = currBox$\\
        $currBox = toRevisit.top()$\\
        $toRevisit.pop()$\\
        $visitedBoxes.insert(currBox)$\\
        $toRevisit.push(currBox)$\\
        \textsc{findNextBox()}\\
    }{
        \textsc{restartOnBoundaryBox()}\\
    }
    \Return

\end{algorithm}

\begin{algorithm}
\caption{Iterator-RestartOnBoundary}
\label{alg::iter::restart}
\SetAlgoLined

    $removeVisited(boundaryBoxes, visited)$\\
    \eIf{$boundaryBoxes$ is not empty}{
        $prevBox = null$\\
        $currBox = boundaryBoxes[0]$\\
        $visitedBoxes.insert(currBox)$\\
        $toRevisit.push(currBox)$\\
    }{
        $isValid = false$\\
    }
    \Return

\end{algorithm}

\begin{algorithm}
\caption{Iterator-Initialize}
\label{alg::iter::init}
\SetAlgoLined

    \SetKwInOut{Input}{input}
    \Input{$tet, plane$}
    $isValid \gets true$\\
    $toRevisit \gets \emptyset$\\
    $visitedBoxes \gets \emptyset$\\
    $boundaryBoxes \gets \textsc{findBoundaryBoxes}(tet)$\\
    $prevBox \gets null$\\
    $currBox \gets boundaryBoxes[0]$\\
    $T \gets tet$\\
    $P \gets plane$\\
    \Return

\end{algorithm}










\end{document}